\documentclass[pra,twocolumn,showpacs]{revtex4}
\usepackage{graphicx}
\usepackage{subfigure}
\usepackage{amsmath,bm,amssymb,}
\usepackage{color}
\makeatletter

\newcommand{\inn}[2]{\vec{#1}\cdot\!\vec{#2}}

\newcommand{\ds}{\displaystyle}

\def \v#1{\vec{#1}}

\newcommand{\imp}{\mathrm{i}}

\newcommand \be {\begin{equation}}
\newcommand \ee {\end{equation}}

\newcommand{\Exp}[1]{\,\mathrm{e}^{\mbox{\footnotesize$#1$}}}
\newcommand{\I}{\mathrm{i}}

\newcommand{\ket}[1]{|#1\rangle}
\newcommand{\bra}[1]{\langle#1|}

\begin{document}

\title{Creation of excitations from a uniform impurity motion in the condensate}
\author{Jun Suzuki}  
\affiliation{Graduate School of Information Systems, The University of Electro-Communications, Tokyo 182-8585 Japan}
\date{\today}


\begin{abstract}
We investigate a phenomenon of creation of excitations in the homogenous Bose-Einstein 
condensate due to an impurity moving with a constant velocity. 
A simple model is considered to take into account dynamical effects due to motions of the impurity. 
Based on this model, we show that there can be a finite amount of excitations created 
even if velocity of the impurity is below Landau's critical velocity. 
We also show that the  total number of excitations scales differently 
for large time across the speed of sound.  
Thus, our result dictates the critical behavior across Landau's one and 
validates Landau's institution to the problem. 
We discuss how Landau's critical velocity emerges and its validity within our model.  
\end{abstract}

\maketitle

\section{Introduction}
Superfluidity has been one of fascinating phenomena in condensed matter physics, and 
has still remained as the central subject of research since its discovery. 
There exist various definitions and criteria whether given a system exhibits 
superfluidity or not, see for example Ref.~\cite{kb2009}. 

It was Landau who gave a major breakthrough to understand this phenomenon based on 
the phenomenological theory, known as the two-fluid model. 
In his seminal paper in 1941, he also gave a simple yet powerful argument 
to determine a critical speed below which a fluid can move without any dissipation \cite{landau41,LLstat2}. 
This critical speed, called Landau's critical velocity, is defined by
\begin{equation}
  \label{eq:1}
v_{{\rm critical}}=\min_{\vec p}\left(\frac{\epsilon(\vec p)}{|\vec p |}\right), 
\end{equation}
where $\epsilon(\vec p)$ is an excitation spectrum of the fluid, and ${\vec p}$ is 
the momentum of excitations in the fluid. When this critical velocity is 
non-zero, the fluid supports dissipation-less motion and hence, the phenomenon of superfluidity occurs. 
As an example consider a fluid whose excitation spectrum is linear in the momentum $p=|\vec{p}|$, 
i.e., $\epsilon(\vec p)=c_s p$, the above critical velocity is equal to the speed of sound 
of the fluid $c_s$. 

The above critical velocity was originally obtained for relative motion of a fluid with 
respect to a container. Similar argument holds when considering a motion of an impurity (an obstacle) 
in the fluid and one can show that if the impurity moves slower than 
Landau's critical velocity, the fluid cannot be excited, see for example, Ref.~\cite{feynman}. 
This implies that an impurity in a fluid can move without friction 
if its velocity is below Landau's critical velocity, thus phenomenon of superfluidity. 
In fact, two kinds of forces to show the superfluidity need to be distinguished: 
One is a drag force onto the fluid when it moves against the impurity, 
and the other is a force acting on the impurity from the fluid.  
These two forces may or may not be same depending on details of 
a given model, i.e., a motion of the impurity, a coupling between the fluid and the impurity, and so on. 

Other aspect of Landau's critical velocity is to infer the stability of a fluid 
as Landau discussed in his paper originally. 
Above this critical velocity, there can be dissipation resulting in a unstable configuration for the fluid. 
This instability is usually referred to as Landau's instability. 
There exists more refined instability criterion from analysis of the solution to non-linear 
partial different equations. Recently, Kato and Watabe proposed 
a unified criterion for the stability of a fluid based on the scaling behaviors of 
the autocorrelation function of the local density \cite{kw2010,watabe10}. 
As will be discussed later, some of our results agree with theirs, 
whereas we are asking a different question and are examining a different physical quantity.  

It is important to remind ourselves the following fact regarding Landau's argument: 
That is his derivation is purely \textit{kinematical} and \textit{classical}. 
There is no guarantee that 
we can apply it to understand dynamical and quantum aspects of this phenomenon.  
Another remark is that the Galilei invariance is a crucial assumption for the 
derivation and one cannot a priori apply this criterion for imhomogeneous systems. 

Bose-Einstein condensates (BECs) are expected to show a phenomenon of superfluidity 
and there have been experimental efforts to examine superfluidity of BECs \cite{becExps1,becExps2,becExps3,becExps4,becExps5}. 
However, the results were rather surprising showing the critical speed was much below 
Landau's one. There were many theoretical analyses on this issue in past to 
understand this discrepancy \cite{becTheo1,becTheo2,becTheo3,fedichev,montina,mazets}. 
Due to the experimental limitation, 
indeed some of assumptions in Landau's derivation are 
not satisfied and thus one needs to analyze many-body problem directly to 
calculate real critical velocity if it exists. 
Drag forces onto BECs when an impurity is immersed were also studied in Refs.~\cite{rp2005,robert05,sdr2009,ccb2012}.
In Ref.~\cite{rp2005}, it was concluded 
that there could be a finite drag force onto BECs at arbitrary small velocity \cite{rp2005,robert05,sdr2009,ccb2012}. 
If this statement is true, Landau's criterion is incorrect and 
BECs cannot show superfluidity in the traditional sense explained in textbooks. 

It is our main motivation here to examine in which sense and under what conditions 
Landau's criterion becomes meaningful and one can use it to check the onset of superfluidity. 
Through this analysis we also wish to resolve some of disagreements in the previous studies. 
For this purpose, we ask a simple, yet a fundamental question: 
How many Bogoliubov excitations (bogolons) are 
created from a single impurity moving with a given constant speed $v$ in BECs. 
In this paper, we analyze a point-like impurity immersed in the homogeneous 
BEC with spatial dimension three. To concentrate on the effect of impurity only, 
we consider the system of the condensate with the impurity at zero temperature in the thermodynamic limit. 
Since this is the situation where impurity version of Landau's criterion seems to hold, 
one expects: There cannot be any excitations created in BEC for $v<c_s$ 
where $c_s$ is the speed of sound of the BEC. 
As will be analyzed in this paper, this naive intuition is incorrect 
and finite amount of bogolons are created at all velocity below the speed of sound. 
However, we also show that its scaling behavior in time 
completely changes across Landau's critical velocity. 
In this way, we give a different interpretation of Landau's critical velocity for the BEC, 
which was not addressed before. 

This paper is organized as follows. 
Sec.~\ref{sec:2} summarizes the model of impurity motion in the BEC and its solution. 
We then analyze asymptotic behaviors of excitations created in the BEC in Sec.~\ref{sec:3} 
and comparisons with numerical analyses are also discussed. 
Sec.~\ref{sec:4} discusses and compares our results with previously known results.  
We close the paper with brief summary and outlook in Sec.~\ref{sec:5}. 

\section{Impurity model and creation of Bogoliubov excitations} \label{sec:2}
We consider a weakly interacting homogeneous condensate at zero temperature. 
The system is described by the Bogoliubov Hamiltonian \cite{B47}: 
\begin{multline} \label{eq:2}
H_{\mathrm{B}}= \int\!d^3x\;\hat{\psi}^{\dagger} (\v{ x},t) (-\frac{\hbar^2
\v{ \nabla} ^2}{2 M})\hat{\psi} (\v{ x},t) \\
+ \frac g2 \int\!d^3x\;\hat{\psi}^{\dagger} (\v{ x},t) \hat{\psi}^{\dagger}
(\v{x},t)\hat{\psi} (\v{ x},t) \hat{\psi} (\v{x},t) , 
\end{multline}
where $M$ is the mass of bosons, $g$ is the coupling constant between bosons, 
and $\hat{\psi} (\v{ x},t)$ is the free boson field operator satisfying the equal-time 
canonical commutation relationship. 
To describe effects of an impurity in BEC, we analyze the following interaction Hamiltonian, 
\begin{equation}
H_{\mathrm{int}}=g_\imp  \int d^3x\;  \rho_\imp(\v x-\v{\zeta}(t))\hat{\psi}^{\dagger} (\v{ x},t) \hat{\psi} (\v{ x},t).  \label{eq:3}
\end{equation}
Here $g_{\imp}$ is the coupling constant between bosons and the impurity, 
$ \rho_\imp(\v x )$ is the density of the impurity at position $\v x$, 
and $\v{ \zeta}(t)$ is a given trajectory for the impurity. 
This impurity model cooperates the effect of local density-density 
interaction between the massive bosons and the impurity. 
For a point-like impurity, we use an approximation 
$\rho_\imp(\v x )=\delta({\v x})$ to simplify the result. 
The effect of impurity size will be discussed in Sec.~\ref{sec:4}. 

In the presence of condensation in the ground state, 
the total Hamiltonian, $H=H_B+H_{\mathrm{int}}$ 
can be simplified using the standard Bogoliubov 
transformation and truncation approximation. 
After simple manipulation, we obtain the effective Hamiltonian 
in the large $N$ and $V$ limit with the fixed density $n=N/V$ as  
\begin{multline} \label{eq:4}
H\simeq  E_0+\sum_{\v k} \hbar \omega _{k} \hat{b}_{\v k}^{\dagger}  \hat{b}_{\v k}\\
+\sum_{\v k} \frac{ng_\imp}{\sqrt{N}}\sqrt{\frac{\epsilon_k}{\hbar\omega_k}}
 \left(\rho_\imp(\v k,t) \hat{b}_{\v k}^{\dagger}+ \rho_\imp(\v k, t)^{*} \hat{b}_{\v k}\right), 
\end{multline}
where neglected terms are smaller than $N^{-1}$. 
In the above expression, $\hat{b}_{\v k}^{\dagger}$ and $\hat{b}_{\v k}$ are the creation and annihilation 
operator describing Bogoliubov excitations (bogolons) without impurity, 
$\epsilon_k=(\hbar k)^2/2M$ is the dispersion relation for free bosons, 
$\omega _{k} = k c_s \sqrt{1+(k\xi)^2}$ is the excitation frequency for the bogolon, 
and $\rho_\imp(\v k,t) $ is the spatial Fourier transformation of the impurity density, 
i.e., $\rho_\imp(\v x-\v{\zeta}(t))$. 
The constants, $c_s=\sqrt{gn/M}$ and $\xi=\hbar/(2Mc_s)$ are the speed of sound 
and the coherent length for the BEC without impurity, respectively, 
and $E_0$ describes the ground state energy without impurity. 
Thus Landau's critical velocity defined by Eq.~\eqref{eq:1} is equal to the speed of sound $c_s$. 
The homogeneous condensate corresponds to the Fock vacuum for 
the bogolons, 
\begin{equation}\label{eq:bec}
\hat{b}_{\v k} \ket{\mathrm{bec}}=0\quad \forall k\neq 0. 
\end{equation}

This impurity Hamiltonian can be diagonalized for any given impurity trajectories
by using a time-dependent unitary transformation, 
and we can evaluate observable quantities \cite{jun05}. 
Let us denote the creation and annihilation operators 
diagonalizing the Hamiltonian \eqref{eq:4} 
by $\hat{c}_{\v k}^{\dagger}$ and $\hat{c}_{\v k}$, respectively. 
Then the effective Hamiltonian reads 
$H\simeq  E'_0+\sum \hbar \omega _{k} \hat{c}_{\v k}^{\dagger}  \hat{c}_{\v k}$. 
These $\hat{c}_{\v k}^{\dagger}(t)$ and $\hat{c}_{\v k}(t)$ describe excitations in the BEC 
dressed by the impurity motion. 
One of fundamental quantity is the expectation value of the occupation number 
in mode $\v k$ at later time $t$ with respect to the homogeneous condensate, i.e., 
\begin{equation}\label{eq:def}
n_{\v k}(t)=\bra{\mathrm{bec}} \hat{c}_{\v k}^{\dagger}  \hat{c}_{\v k}\ket{\mathrm{bec}}. 
\end{equation}
One can show that this number is same as the expectation value of $ \hat{b}_{\v k}^{\dagger}  \hat{b}_{\v k}$ 
with respect to the Fock vacuum for $\hat{c}_{\v k}$. 
The occupation number is calculated by assuming that the system is in the ground state, i.e., the homogeneous BEC, 
at initial time $t_0$ as 
\begin{align} \label{eq:4-1}
n_{\v k}(t)& = \frac{ng_\imp^2}{V\hbar^2}\,\frac{\epsilon_k}{\hbar\omega_k}\, |I_{\v k}(t)|^2, \\
 I_{\v k}(t)&=\int_{t_0}^t dt' \rho_\imp(\v k,t') \Exp{\I \omega_k t'} .  \label{eq:4-2}
\end{align}

As stated in Introduction, we focus on a creation process 
of bogolons from an impurity moving with a constant speed $v$ along a fixed direction, say $z$-axis. 
The trajectory is $\vec\zeta (t)=(0,0,vt)$ for $t\ge t_0$ at which the impurity starts to move.  
We analyze time-dependence of bogolons emitted in the BEC. 
For this case, $ \rho_\imp(\v k,t)=\exp(-\I \inn{k}{v}t)$ holds 
and we can carry out the time integral in \eqref{eq:4-2}. 
The number of bogolons created in a solid angle $d\Omega$ within a time interval $t$, i.e., 
later time measured from $t_0$, 
is calculated in the thermodynamic limit (replacing the summation by $k$ integral) as 
\begin{align}\nonumber
\frac{d{\cal N}(v,t)}{d\Omega}
&=\int_0^\infty\frac{dk}{(2\pi)^3}\;k^2 \,V n_{\v k}(t)\\
&=\frac{2ng_\imp^2}{(2\pi)^3\hbar^2}\int_0^{\infty}\! dk\; k^2\frac{\epsilon_k}{\hbar\omega_k}\,
\frac{1-\cos[(\omega_k -\inn{k}{v})t]}{(\omega_k -\inn{k}{ v})^2}, \label{eq:5}
\end{align}
with $\v v=(0,0,v)$ the velocity of the impurity. 
The total number of created bogolons at time $t$ is then obtained by integrating 
over the angle variables as 
\be \label{eq:6}
{\cal N}(v,t)=\int d\Omega\ \frac{d{\cal N}(v,t)}{d\Omega}, 
\ee
which is equivalent to $\int d^3k/(2\pi)^3\;V n_{\v k}(t)$. 
This number ${\cal N}(v,t)$ counts the total number of excitations created in the BEC 
due to the impurity and this is the quantity of our main interest in the rest of paper. 

Before passing to the next section, we have several remarks. 
First, the initial condition is such that impurity was absent before time $t_0$. 
This is contrast to the situation where the interaction Hamiltonian is adiabatically 
switched off with the standard adiabatic factor $\exp(\eta t)$, where a positive parameter $\eta$ goes to $0$ at 
the end of calculations. 
Second, the total occupation number \eqref{eq:6} is well-defined for 
the singular point $\omega_k -\inn{k}{ v}=0$ in the denominator of the integrand. 
This integral also converges for short wave length and thus no subtraction is needed. 
Thus, this quantity is easily evaluated numerically without any artificial cut-offs.  
Third, this creation process is not obtained within the linear response thoery, 
since ${\cal N}(v,t)$ is proportional to the square of the coupling constant $g_\imp$. 
We give analysis of effects within the standard linear response theory in Sec.~\ref{sec:4}. 
Last, from the above integrals (\ref{eq:5},\ref{eq:6}) one does not expect 
these integrals to vanish for a static impurity ($v=0$). 
This means that a point-like obstacle pinned at some position can 
excite the BEC without any motion. We thus need to distinguish this effect, referred 
to as the static effect from dynamical effects due to the motion of impurity 
in the following discussion. This matter will be discussed in Sec.~\ref{sec:stat}

\section{Results} \label{sec:3}
In this section, we give detail analysis on the integral  (\ref{eq:5},\ref{eq:6}) for large time $t$ limit.  
In the following we measure time and length in units of the speed of sound and 
the coherence length to simplify notations unless noted explicitly. 
That is, time and length are in units of $\xi/c_s$ and $\xi$, respectively.  
The speed of impurity is normalized by $c_s$ and we use the notation $\beta=v/c_s$.  
With this convention, for example, the integral \eqref{eq:5} reads
\begin{multline} \label{eq:5-1}
\frac{d{\cal N}(\beta,t )}{d\Omega}
=\\
\frac{{\cal N}_0}{2\pi}\int_0^{\infty}\!\! dk \; \frac{k ^3}{\sqrt{k ^2\!+\!1}}\,
\frac{1-\cos[(k \sqrt{k ^2\!+\!1} -\vec{k}\!\cdot\! \vec{v})t ]}
{(k \sqrt{k ^2\!+\!1} -\vec{k}\!\cdot\! \vec{v})^2},
\end{multline}
with ${\cal N}_0$ a dimensionless constant defined by
\begin{equation} \label{defN0}
{\cal N}_0=\frac{ng_i^2}{8\pi^2 \xi^3(Mc_s^2)^2}. 
\end{equation}

As we see from the above expression, the integrand decays for large $k $ 
while the numerator oscillating in $k $. The asymptote of this integral 
is examined by applying the method of stationary phase with singular boundary conditions \cite{erdelyi}. 
Before going to asymptotic analysis, we first evaluate infinite time limit 
reproducing the previously known result. One of main interest is to see how 
finite time result merges to the infinite time limit. 
Following the infinite time limit analysis, 
we give thorough asymptotic analysis together with numerical analysis. 

\subsection{Infinite time limit} \label{inftime}
When the impurity moves from the initial time $t_0$ to infinity remote future, we can use 
the formula $\lim_{t\to\infty}(1-\cos \omega t)/\omega^2t=\pi\delta(\omega)$ 
for the integral \eqref{eq:5} to get
\begin{align} \nonumber \label{eq:7}
\lim_{t  \to\infty}\frac{1}{t } \frac{d{\cal N}(\beta,t )}{d\Omega}
&=\frac{{\cal N}_0}{2}\int_0^{\infty} dk \;\frac{k ^3}{\sqrt{k ^2+1}}\,\\ \nonumber
&\hspace{1.2cm}\times\delta \left(k \sqrt{k ^2+1}-k  \beta\cos\theta\right)\\
&=\frac{{\cal N}_0}{2}\sqrt{(\beta\cos\theta)^2-1}\;\Theta(v\cos\theta-1), 
\end{align}
where $\theta$ is the angle between $\v k$ vector and $z$-axis 
and $\Theta(x)$ is the Heviside step function. 
This limit, the left hand side of Eq.~\eqref{eq:7}, represents 
the average number of bogolons created in infinite time limit within the solid angle $d\Omega$, 
or it is the steady-state excitation rate at angle $d\Omega$ \cite{comment1}. 
Owing to the Heviside step function $\Theta(\beta\cos\theta-1)$, 
bologons can be created if and only if 
$\beta\ge1\Leftrightarrow v\ge c_s$ and in this case 
the emission occurs only within the forward cone; $|\theta|\le\theta_c=\cos^{-1}(\beta)$.  
Another important consequence for the case $\beta>1$ is that existence of 
cut-off given by $k_c=\sqrt{\beta^2-1}$ above which no excitations are possible. 

The total average number of bogolons, i.e., the steady-state excitation rate, 
is then obtained by integrating over angles as
\begin{multline} \label{eq:8}
\lim_{t  \to\infty}\frac{1}{t } {\cal N}(\beta,t )\\
=\frac{\pi{\cal N}_0}{2\beta}\big[\beta\sqrt{\beta^2-1}-\log(\beta+\sqrt{\beta^2-1})\big]\Theta(\beta-1).
\end{multline}
This result shows that the average number of bogolons created due to the impurity 
is non-zero if and only if the speed of impurity exceeds the speed of sound $c_s$. 
An intuitive argument for this result is that, 
in this infinitely time limit, no dynamics plays important role and hence 
only kinematically permitted processes can happen. 
From this expression \eqref{eq:8}, we see that 
$\lim_{t  \to\infty}{\cal N}(\beta,t )/({\cal N}_0t )$ 
behaves as $(\pi\sqrt{8}/3)(\beta-1)^{3/2}$ for $\beta$ sufficiently close to $1$, i.e., $0<\beta-1\ll 1$. 
Thus Eq.~\eqref{eq:8} and tis first derivative approach to zero as $\beta\to1$ from above,
whereas higher derivatives diverge as $\beta\to1$. 

We note that this result was obtained by several authors for different models and different approximations. 
The easiest way to obtain it is to calculate the decay rate using the Fermi's golden rule within 
second order approximation in the interaction Hamiltonian \cite{feynman}. 
Physical mechanism of this process is in fact same as the Cherenkov radiation known in 
quantum electrodynamics; a charge particle moving with a uniform velocity 
creates radiation when its speed exceeds an effective speed of light in some medium. 
This ``Cherenkov radiation" in the homogeneous BEC was also studied 
previously, for example \cite{kovrizhin,astracharchik}. 
Lastly, we comment that we are not examining radiation but the number of created bogolons. 
Two quantities are, of course, related to each other, yet they provide 
quantitatively different behaviors in general \cite{comment0}. 

\subsection{Asymptotic analysis}
We now analyze the integral \eqref{eq:5-1} for large $t$. 
For this purpose, we define the following integral:
\begin{align}\nonumber
I(\upsilon,\tau)&=
\int_0^{\infty}\!\! dk\; \frac{k}{\sqrt{k^2+1}}\,
\frac{1-\cos[k(\sqrt{k^2+1} -\upsilon)\tau]}
{(\sqrt{k^2+1} -\upsilon)^2}\\  \label{eq:int}
&=\int_1^{\infty}\!\! d\kappa\; \,
\frac{1-\cos[\sqrt{\kappa^2-1}(\kappa -\upsilon)\tau]}
{(\kappa -\upsilon)^2},
\end{align}
where the second integral representation is obtained 
with the change of variable $\kappa=\sqrt{k^2+1}$. 
With this definition, the anglar distribution of created bogolons 
is expressed simply as 
\be\label{eq:5-2}
\frac{d{\cal N}(\beta,t )}{d\Omega}=\frac{{\cal N}_0}{2\pi}\;I(\beta\cos\theta,t ). 
\ee
We analyze large $\tau$ asymptotic behaviors of the integral \eqref{eq:int} 
for a given value of $\upsilon$. Define the following function appeared 
in argument of cosine function by
\be\label{eq:9}
\phi_\upsilon(\kappa)=\sqrt{\kappa^2-1}(\kappa -\upsilon),
\ee
then asymptote of the integral \eqref{eq:int} depends on 
whether or not  the denominator of the integrand and 
the derivative of Eq.~\eqref{eq:9} vanish during the integration over $\kappa$ \cite{erdelyi}. 

Since the first order derivative of $\phi_\upsilon(\kappa)$ is 
\be\label{eq:10}
\phi_\upsilon'(\kappa)=\frac{d}{d\kappa}\phi_\upsilon(\kappa)
=\frac{2 \kappa^2-\upsilon \kappa-1}{\sqrt{\kappa^2-1}},
\ee
we observe the following three cases 
need to be analyzed separately; (a) $\upsilon<1$, (b) $\upsilon=1$, and (c) $\upsilon>1$. 

\subsubsection{Case (a): $\upsilon<1$}
The denominator of \eqref{eq:int} and the first order derivative do not 
vanish for all values $\kappa\ge1$. 
Then the integration by parts gives a proper asymptotic expansion of the integral 
and the result is 
\be \label{eq:11}
I(\upsilon,\tau)=(1-\upsilon)^{-1}+(1-\upsilon)^{-4}\tau^{-2}+{\cal O}(\tau^{-4}). 
\ee
Here, neglected terms are expressed as the largest order by ${\cal O}(\tau^{-s})$. 

\subsubsection{Case (b): $\upsilon=1$}
This case is critical in some sense from the asymptotic analysis point of view. 
This is because the denominator of the integrand and 
the derivative Eq.~\eqref{eq:10} vanish at the integration boundary $\kappa=1$. 
With a refined version of the method of stationary phase, the major contribution 
is due to the boundary $\kappa=1\Leftrightarrow k=0$, i.e., zero mode. The result is
\be \label{eq:12}
I(\upsilon,\tau)=2^{-2/3}\Gamma(1/3)\tau^{2/3}+{\cal O}(\tau^{-1/3}), 
\ee
where $\Gamma(x)$ denotes the gamma function. 
Thus the integral $I(\upsilon,\tau)$ diverges as $\tau$ approaches to infinite. 

\subsubsection{Case (c): $\upsilon>1$}
The last case needs further care due to appearance of several singularities 
in the derivative Eq.~\eqref{eq:10}. Let $\kappa_{\pm}$ be the solution 
to $2 \kappa^2-\upsilon \kappa-1=0$ with a convention 
$\kappa_+ >\kappa_-$ 
(The relation $\upsilon>\kappa_+$ holds in this case.), 
and divide the integration interval of \eqref{eq:int} into three parts as
\be \label{eq:13}
I_j(\upsilon,\tau)=\int_{D_j}\!\! d\kappa\; \,
\frac{1-\cos[\phi_\upsilon(\kappa)\tau]}
{(\kappa -\upsilon)^2},
\ee
with 
\begin{equation}
D_1=[1,\kappa_+],\;
D_2=[\kappa_+,\upsilon],\;
D_3=[\upsilon,\infty). \label{eq:14} 
\end{equation}

Asymptotes of the integrals $I_j(\upsilon,\tau)$ can be evaluated separately as follows. 
Firstly, we have
\begin{multline}\label{eq:15}
I_1(\upsilon,\tau)=(\upsilon-\kappa_+)^{-1}-(\upsilon-1)^{-1}\\
-(\upsilon-\kappa_+)^{-2}\sqrt{\frac{\pi}{2\phi''_\upsilon(\kappa_+)}}
\frac{\cos[\phi_\upsilon(\kappa_+)\tau+\pi/4]}{\tau^{1/2}}\\
+{\cal O}(\tau^{-1}).
\end{multline}
The major contribution to the integral here is the boundary $\kappa=\kappa_+$ 
and the other boundary $\kappa=1$ gives terms of order $\tau^{-5/3}$. 
We remind $\kappa=\kappa_+\Leftrightarrow k =\sqrt{\upsilon^2-4+\upsilon\sqrt{\upsilon^2+8}}/\sqrt{8}$ 
corresponds to the stationary point of the integral, i.e., $\phi'_\upsilon(\kappa_+)=0$. 

Secondly, we obtain
\begin{multline}\label{eq:16}
I_2(\upsilon,\tau)=-(\upsilon-\kappa_+)^{-1}\\
-(\upsilon-\kappa_+)^{-2}\sqrt{\frac{\pi}{2\phi''_\upsilon(\kappa_+)}}
\frac{\cos[\phi_\upsilon(\kappa_+)\tau+\pi/4]}{\tau^{1/2}}\\
+\frac{\pi}{2}\sqrt{\upsilon^2-1}\;\tau+{\cal O}(\tau^{-1}).
\end{multline}
The term in the second line is due to the boundary at $\kappa=\kappa_+$ and 
the first term of the last line is due to $\kappa=\upsilon$. 
Apparently, the latter term dominates the integral and 
the stationary point does not contributes significantly. 

Lastly, we have 
\begin{equation}\label{eq:17}
I_3(\upsilon,\tau)=
\frac{\pi}{2}\sqrt{\upsilon^2-1}\;\tau
{+{\cal O}(\tau^{-1})}.
\end{equation}

Adding three integrals we obtain asymptote for the case (c) as
\be\label{eq:18}
I(\upsilon,\tau)=-(\upsilon-1)^{-1}+\pi\sqrt{\upsilon^2-1}\;\tau{+{\cal O}(\tau^{-1})}.
\ee

\subsection{Creation of bogolons}
\subsubsection{Anglar distribution of created bogolons}
Angular distribution of created bogolons at large time $t\gg1$ is thus obtained 
depending upon the value of $\beta\cos\theta$ as
\be\label{eq:19}
\frac{2\pi}{{\cal N}_0}\frac{d{\cal N}(\beta,t )}{d\Omega}=
\begin{cases}
\ds(1-\beta\cos\theta)^{-1}+(1-\beta\cos\theta)^{-4}\;t ^{-2}\\
\hspace{3.2cm}\mathrm{for}\ \beta\cos\theta<1\\[2mm]
\ds2^{-2/3}\Gamma(1/3)\;t ^{2/3}
\hspace{0.65cm}\mathrm{for}\ \beta\cos\theta=1\\[4mm]
\ds\pi\sqrt{(\beta\cos\theta)^2-1}\;t -(\beta\cos\theta-1)^{-1}\\
\hspace{3.2cm}\mathrm{for}\ \beta\cos\theta>1
\end{cases}.
\ee
From this result, we observe that 
the average number of created bogolons, i.e., $d{\cal N}(\beta,t )/td\Omega$, 
approaches to the infinite time limit \eqref{eq:7}. 
Its convergence is polynomial in time and hence rather slow. In this limit, however, 
one looses different scaling behavior in time across the value $\beta\cos\theta=1$. 
This shows that the infinite time limit $t\to\infty$ does not commute with 
the limit for $\beta\cos\theta$ approaching to $1$. 
We observe that the appearance of clear forward cone as 
predicted by \eqref{eq:7} does not show up in finite time analysis. 
Because of $-1\le\cos\theta\le1$, the speed of impurity needs to exceed 
the speed of sound in order to see different scaling behaviors, 
otherwise one only observes a simple decaying behavior. 

\subsubsection{Total number of created bogolons} \label{sec:asympt}
We analyze the asymptotic behavior of the total number of bogolons created in the homogenous BEC. 
When the speed of impurity $v$ is smaller than $c_s$, we can simply 
integrate the result \eqref{eq:19} over angles. When $\beta\ge1$, 
we need further analysis which is similar to the one used in the previous subsection. In the following 
we only show the result with leading orders without details. The result is
\be\label{eq:20}
\frac{{\cal N}(\beta,t )}{{\cal N}_0}\simeq
\begin{cases}
\ds\frac{1}{\beta}\log(\frac{1+\beta}{1-\beta})-\frac23\frac{\beta^2+2}{(1-\beta^2)^3}\;t ^{-2}\\
\hspace{4.45cm}\mathrm{for}\ \beta<1\\[2mm]
\ds\frac23\gamma+\frac73\log 2+\frac23\log t 
\hspace{1.15cm}\mathrm{for}\ \beta=1\\[4mm]
\ds\frac{\pi}{2\beta}\left[\beta\sqrt{\beta^2-1}-\log(\beta+\sqrt{\beta^2-1})\right]\;t \\
\ds\hspace{1.35cm}+\frac{1}{\beta}\log\big(\frac{\beta+1}{\beta-1}\big)\hspace{7mm}\mathrm{for}\ \beta>1
\end{cases},
\ee
where $\gamma\simeq0.577$ is the Euler's constant. 
This quantity \eqref{eq:20} counts the transient Bogoliubov excitations 
and we see that it approaches to the steady-state rate \eqref{eq:8} 
in the limit $\lim_{t\to\infty} {\cal N}(\beta,t)/t$ for any given value of impurity speed $\beta$. 

As we can see from the expressions \eqref{eq:19} and \eqref{eq:20}, 
divergent behaviors occur 
when $\beta\cos\theta$ and $\beta$ approaches to $1$ from both sides. 
This is due to the fact that we are examining asymptotic behavior in large time $t$ for a given value of $\beta$. 
To see the asymptotic behaviors 
near $\beta=1$, we expand the integral \eqref{eq:6} as a function of $\beta$ around $1$
and then examine its asymptotic behavior of expansion coefficients separately. 
Similar analysis yields for sufficiently small 
$|\beta-1|\ll 1$: 
\begin{multline} \label{eq:21}
\frac{1}{{\cal N}_0}\,{{\cal N}(\beta,t )}=\frac23\gamma+\frac73\log 2+\frac23\log t \\
+\left[\frac12-\frac23\gamma-\frac73\log 2+2^{-2/3}\Gamma(1/3)\;t ^{2/3}\right]\,(\beta-1)\\
{+{\cal O}\big((\beta-1)^{2}\big)}.
\end{multline}
In this way, we observe smooth transition across $\beta=1$. 
This matter near $\beta=1$ is also analyzed numerically in the next subsection. 

\subsubsection{The static effect} \label{sec:stat}
As mentioned in remarks of the last section, we analyze the static effect 
due to a pinned point-like impurity in the BEC. 
Setting $v=0$, i.e., ${\cal N}_{\mathrm{stat}}(t )={\cal N}(\beta=0,t )$, we have 
\begin{equation} \label{eq:22}
\frac{{\cal N}_{\mathrm{stat}}(t )}{{\cal N}_0}=
2\int_0^{\infty}\!\! dk \; \frac{k }{(k ^2\!+\!1)^{3/2}}\,
\big(1-\cos[k \sqrt{k ^2\!+\!1} t ]\big). 
\end{equation}
Similar asymptotic analysis gives 
\be \label{eq:23}
\frac{{\cal N}_{\mathrm{stat}}(t )}{{\cal N}_0}=2+4t ^{-2}+{\cal O}(t ^{-4}). 
\ee
Thus we can interpret the constant ${\cal N}_0$ defined in Eq.~\eqref{defN0} as 
twice the number of created bogolons when one insert a single impurity in the homogeneous condensate. 
This contribution should be subtracted when one is interested only in effects 
due to motions of impurities. We note that this number can also be interpreted as 
condensate deformation due to an impurity immersed in the homogeneous BEC \cite{gps1994,castin,mg12}.

\subsection{Numerical analysis}
\begin{figure}[htbp]
\begin{center}
\includegraphics[width=8.5cm]{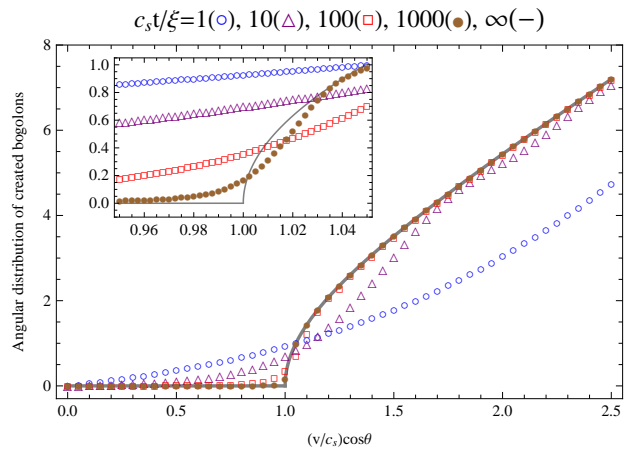}
\caption{The angular dependence of average number of created bogolons \eqref{eq:5} divided by time 
as a function of $\beta \cos\theta$ for four different values of time: 
$c_s t/\xi=1(\circ),10({\scriptstyle\triangle}),100({\scriptstyle\Box}),1000(\bullet)$. 
The grey line represents the infinite time limit \eqref{eq:7}. 
The number of excitations is measured in units of ${\cal N}_0/2\pi$.}
\label{fig1}
\end{center}
\end{figure}

\begin{figure}[htbp]
\begin{center}
\includegraphics[width=8.5cm]{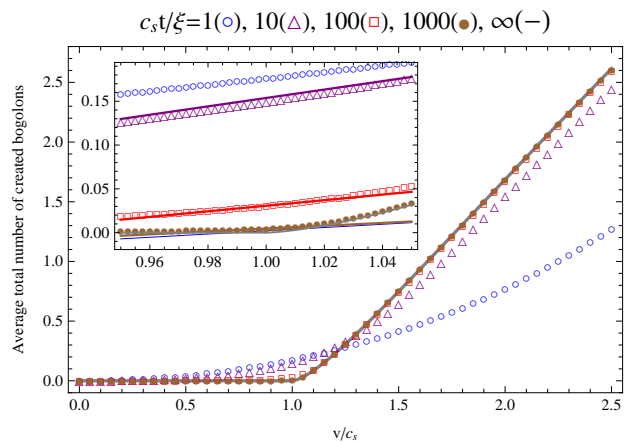}
\caption{The average number of created bogolons \eqref{eq:6} divided by time 
as a function of $\beta=v/c_s$ for four different values of time: 
$c_s t/\xi=1(\circ),10({\scriptstyle\triangle}),100({\scriptstyle\Box}),1000(\bullet)$. 
The grey line represents the infinite time limit \eqref{eq:8}. 
The number of excitations is measured in units of ${\cal N}_0$. 
In the inset, the solid lines with corresponding colors show asymptotic result given by Eq.~\eqref{eq:21}.}
\label{fig2}
\end{center}
\end{figure}

We numerically evaluate the integrals (\ref{eq:5},\ref{eq:6}) for several values $\beta$ and $t $. 
Here we have two choices of plotting figures for created bogolons. 
One is to plot average numbers of created bogolons, as a function of speed of the impurity for a given value of time, 
and the other is as a function of time for a given impurity speed. 
The former manner of plotting has been used in the previous studies, 
yet we claim in this paper that this is not a proper way to analyze the critical behavior in $\beta$. 
In this case, as shown in FIGs.~\ref{fig1}, \ref{fig2}, 
one can only observe smooth curves around $\beta=1$ for any finite values of $t $. 
The latter manner, on the other hand, captures the critical behavior across $\beta=1$ 
in the large time regime. To make our discussion clear, we present these two methods separately. 
For all figures, we subtract the static effect ($\beta=0$) as mentioned in \ref{sec:stat}. 

FIG.~\ref{fig1} shows the average number of bogolons created as a function of $\beta \cos\theta$, 
i.e., the integral \eqref{eq:5} divided by time $t $, for four different values of time: 
$c_s t/\xi=1,10,100,1000$. The number of excitations is measured in units of ${\cal N}_0/2\pi$. 
To compare it with the ideal case, i.e., infinite time limit, we also plot the result \eqref{eq:7} which 
is shown by the grey line. The division by time and hence to plot the average number 
is important in order to compare the result with the infinite time limit case.

It is clear from FIG.~\ref{fig1} that one does not see non-analytic behavior any values of finite time. 
This is seen clearly in the inset of FIG.~\ref{fig1}. 

In FIG.~\ref{fig2} we plot the average total number of created bogolons as a function of $\beta$, 
for four different values of time: 
$c_s t/\xi=1,10,100,1000$. The number of excitations created 
in the BEC is measured in units of ${\cal N}_0$ here. 
We plot the ideal case \eqref{eq:8} for a comparison and this is shown by the grey line. 
FIG.~\ref{fig2} shows the dependence on the impurity speed from $0$ to $2.5$, 
and the inset does around the value $\beta=1$. 
Although we see large time cases approach to the infinite time limit in FIG.~\ref{fig2}, 
one never get non-analytic behaviors. 
This matter can be seen more clearly in the inset of FIG.~\ref{fig2}. 
To compare our asymptotic result \eqref{eq:21} with the numerical analysis, 
asymptotes are plotted in the inset showing good agreements except for $t =1$ 
which is not to be regarded as large value. 

\begin{figure}[htbp]
\begin{center}
\includegraphics[width=8.5cm]{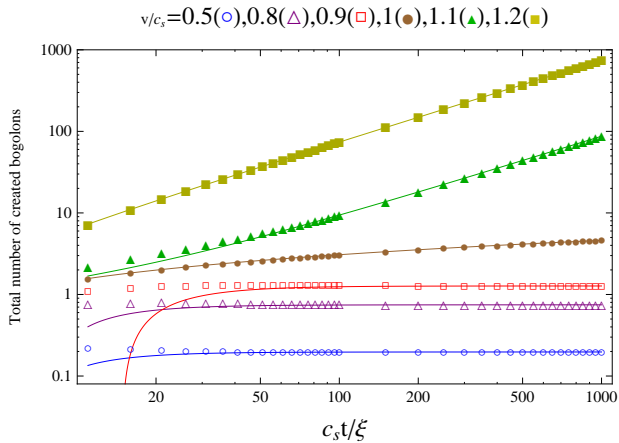}
\caption{The total number of created bogolons \eqref{eq:5} as a function of 
time $t $ for six different values of impurity speed:
$\beta=v/c_s=0.5({\circ}), 0.80({\scriptstyle\triangle}), 0.9({\scriptstyle\Box}),1({\bullet}),
1.1({\scriptstyle\blacktriangle}), 1.2({\scriptstyle\blacksquare})$. 
Asymptotic analysis for large $t$ \eqref{eq:20} are shown by solid curves with corresponding colors.
The number of excitations is measured in units of ${\cal N}_0$.}
\label{fig3}
\end{center}
\end{figure}

FIG.~\ref{fig3} shows the numerical analysis of the total number of 
created bogolons as a function of time for six different values of impurity speed: 
$\beta=v/c_s=0.5, 0.8, 0.9,1,1.1, 1.2$. 
Excitations are measured in units of ${\cal N}_0$ and both axes are shown in logarithmic scale. 
Asymptotic results \eqref{eq:20} are also shown as solid curves. 
From this result we observe validity of asymptotes obtained in Sec.~\ref{sec:asympt}, 
in particular Eq.~\eqref{eq:20}. Importantly, we numerically confirm that scaling in time changes 
completely depending on the value of impurity speed across $\beta=1$, 
that is from constant, logarithmic, and then to linear. 
This different scaling behaviors confirm the criticality and 
Landau's critical velocity $v=c_s$ is indeed \textit{critical} in this sense. 

\begin{figure}[htbp]
\begin{center}
\includegraphics[width=8.5cm]{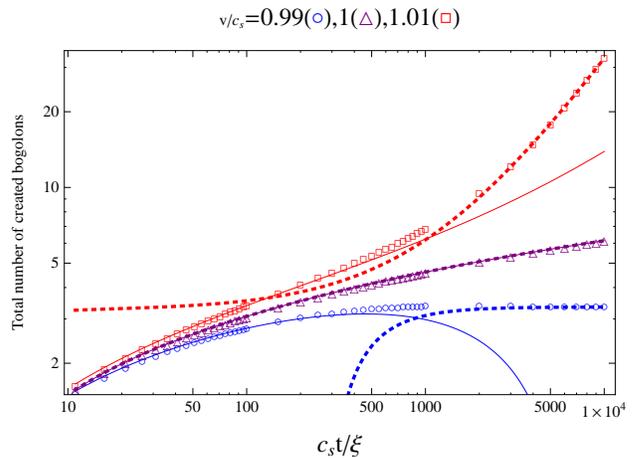}
\caption{The total number of created bogolons \eqref{eq:5} as a function of 
time $t $ near critical velocity $\beta=1$ for three different values of impurity speed:
$\beta=v/c_s=0.99(\circ), 1({\scriptstyle\triangle}),1.01({\scriptstyle\Box})$. 
Asymptotes \eqref{eq:20} are shown by dotted curves and asymptotic analysis around 
$v=c_s$ \eqref{eq:21} are shown by solid curves. 
The number of excitations is measured in units of ${\cal N}_0$.}
\label{fig4}
\end{center}
\end{figure}

To see asymptotic behavior near the critical velocity $\beta=1$, we 
analyze the total number of created bogolons for $\beta=v/c_s=0.99, 1, 1.01$ in FIG.~\ref{fig4}. 
There, we show asymptotes \eqref{eq:20} by dotted curves and asymptotic analysis around 
$\beta=1$ \eqref{eq:21} by solid curves. This figure confirms that 
our asymptotic analysis is correct as time $c_s t/\xi$ becomes large even near the critical velocity. 
FIG.~\ref{fig4} also shows transition in better approximations from Eq.~\eqref{eq:21} to Eq.~\eqref{eq:20} 
as time gets larger. For relatively small time intervals, the asymptote \eqref{eq:21} 
fits quite well and then the asymptote \eqref{eq:20} becomes valid approximation 
for large time intervals as is expected. Similar behaviors can be observed other values 
of impurity speed near $\beta=1$. 

\section{Discussions} \label{sec:4}
In this section, we discuss our results in detail and compare with previous studies. 
\subsection{Further analysis of results}
\subsubsection{Creation of excitations from slow impurity motion}
We analytically and numerically confirm 
that there exist Bogoliubov excitations (bogolons) created 
when an impurity moves in the condensate with 
arbitrary constant velocity $v$ even below the speed of sound $c_s$, 
i.e., Landau's critical velocity. This happens for three-dimensional 
homogeneous condensate in the thermodynamic limit at zero temperature, that is this effect is 
not due to any finite size effects or finite temperature effects. 

The main mechanism of this creation process below Landau's critical velocity 
is that an impurity motion can excite bogolons with low momenta for short time scale. 
Only when one measures the steady-state excitation rate, i.e., 
the average number of created excitations after infinitely long time, 
which is calculated by the number of excitations divided by time, 
one sees that it behaves as Landau predicted originally, 
i.e., the steady-state excitation rate drops to zero below the speed of sound. 
Importantly, corrections from finite time do not vanish exponentially, 
but decay only polynomial in time, rather slow convergence. 
We thus expect that this is an observable effect in trapped BEC systems as well. 

Above the critical velocity, on the other hand, the main contribution 
is not due to the stationary phase point in the integrals (\ref{eq:5},\ref{eq:6}) but due to 
bogolons around the momentum satisfying the condition: $k\xi=[(v/c_s)^2\cos^2\theta-1]^{1/2}$. 
Combining with the inequality $|\cos\theta|\le1$, 
we have a necessary condition $v/c_s>1$ for 
this momentum to lie in the integration interval. Otherwise, 
contribution from low momenta $k\xi\ll 1$ dominates the integral. 

An interesting observation from our analysis is that 
when the impurity speed is below the speed of sound, 
the total number of created bogolons converges to 
a fixed value ${\cal N}_0(v/c_s)^{-1}\log[(c_s+v)/(c_s-v)]$ after 
the impurity moves for sufficiently large time interval. 
This value as a function of impurity speed 
is even function and behaves 
as ${\cal N}_0 [2+(v/c_s)^2/3+{\cal O}((v/c_s)^4)]$ 
for sufficiently small values of $v/c_s$. 
The first term is nothing but the static effect evaluated in \ref{sec:stat}, 
and the second term predicts that creation of bogolons increases 
quadratic in the impurity speed for slow impurity motions.  

A key fact in our results is that two limits do not commute with each other: 
One is $t\to\infty$ and the other is $v\to c_s$ from above and below. 
If one takes the limit $t\to\infty$ first, one may observe non-analytical 
behavior for the average number of crated bogolons as Eq.~\eqref{eq:8}. 
This expression only tells us that there is no bogolons created average in time 
at and below Landau's critical velocity. However, one cannot conclude 
whether or not we have created bogolons in these two cases $v\le c_s$. 
Our finite time analysis shows that bogolons are indeed created 
even below Landau's critical velocity and furthermore 
it grows logarithmically in time at this critical velocity. 

\subsubsection{Validity of assumptions}
Here we discuss some of assumptions made in our analysis. 
First, the effective Hamiltonian \eqref{eq:4} \cite{comment2}. 
The assumption to derive it 
is that a coupling between an impurity and bosons is weak 
so that impurity effect can be treated as a perturbation. 
This effective Hamiltonian becomes good approximation when the total number of bosons 
and the size of the system are sufficiently large. 
However, this large $N$ and $V$ limit 
needs further clarification since the system of the homogenous 
condensate plus an impurity may not be a homogeneous 
system in the thermodynamics limit \cite{zbyszek}. 
Generally speaking, when an impurity immersed in BECs, 
it produces condensate deformation and 
should change some of macroscopic as well as microscopic properties of BECs, 
such as the dispersion relation, speed of sound, and so on \cite{gps1994,castin,mg12}. 
In this paper, we do not include this condensate deformation 
since we are interested only in the response of homogeneous condensate 
against an impurity perturbation. That is to start with the homogenous 
ground state for the unperturbed Hamiltonian. 
This assumption of treating the impurity as a small perturbation is valid 
if we consider weak coupling between the impurity and the BEC. 
Strength of the coupling strongly depends upon a specific choice for 
impurity species and BECs and we can choose a system satisfying 
our assumption. 

Second, large time asymptotic analysis to derive our results. 
With our convention, time is measured by $\xi/c_s$. 
For usual BEC systems, this number is rather small, typically 
on the order of $10^{-5}$ sec. Comparison with numerical 
analysis implies $100 \xi/c_s\sim$ a few m sec is already 
to be considered as large time when
$v$ is not too close to $c_s$. This time scale 
is within a possible range for current experiments 
with trapped BECs and it is in principle possible to 
measure critical behaviors analyzed in this paper. 

Third, the point-like impurity assumption. 
In our analysis we consider an impurity with zero-size limit. 
Size of the impurity in general should give rise to observable effects. 
As stated above, a small impurity creates 
local depletion of condensate and this is more visible as 
the size of impurity becomes bigger. 
Within our assumptions, the present model Hamiltonian 
does not change the critical velocity as discussed in Sec.~\ref{sec:finitesize} below. 
This result also suggests that our treatment needs 
to be refined to take into account these effects properly
when we are interested in the effect due shapes and size of impurities. 

Last, neglect of back reaction onto the impurity. 
This assumption needs to be verified carefully 
if one is interested in observing creation of excitations 
from very slow impurity motions. 
In this case there will be also an effective potential for the impurity 
acting from BEC. It may be necessary to cancel 
a force from this potential by applying additional force on the impurity 
externally. It becomes more difficult to maintain the constant speed for the impurity. 
Thus our result is only valid for small coupling to the impurity,
 yet its speed needs to be not too slow. 
Effects of back reaction onto the impurity have richer physics, 
and we shall report progress on this front in due course. 

\subsubsection{Choice of initial condition} \label{initcond}
Throughout our analysis, we consider a situation where 
an impurity starts to move at finite initial time $t_0$ in the homogeneous condensate. 
This initial condition seems to be very crucial to obtain 
our result. The other possible initial condition is to 
take $t_0\to-\infty$ limit. In this case, however, one 
needs to use a standard mathematical trick, an adiabatic factor, 
given by $\exp(\eta t)$ with $\eta$ a small positive number going to zero 
at the end of calculations. In this way, 
the integral \eqref{eq:4-2} converges in $t_0\to-\infty$ limit. 
With this standard treatment, one can show that the occupation number 
\eqref{eq:4-1} does not have any time-dependence 
and thus no dynamics for bogolons created from a uniform impurity motions. 

This latter choice of initial condition $t_0\to-\infty$ needs to be 
further clarification when one considers a linear motion. 
This is because one needs to have infinite volume size for 
the condensate from the beginning in order for the impurity 
to moves infinitely remote past. Switching off of impurity 
effect adiabatically also needs to be justified. 
Our expectation without any detail calculations is that any impurity motion 
accompanying acceleration creates bogolons and one cannot 
simply represent this process by the adiabatic factor only. 
This is based on our preliminary analysis on radiation 
from periodic motions of an impurity, such as dipole oscillation and circular motion \cite{jun06}. 
Rather we need to examine dynamical effects due to 
the impurity motion with an explicit time-dependent trajectory, 
such as $\zeta(t)=vt [\tanh (\gamma t)+1]/2$ with $\gamma$ a parameter. 
Additionally, we need to check if our result is due to a sudden acceleration 
from some fixed position to a uniform motion. The trajectory in this case 
is given by $\zeta(t)=\zeta_0$($=$constant) for $t_1> t\ge t_0$ and $\zeta(t)=v(t-t_1)+\zeta_0$ for $t\ge t_1$ 
with an additional parameter $t_1$ at which the impurity starts to move. 
Further investigations need to investigate this issue, which is left for future study. 


\subsubsection{Finite impurity size effect}\label{sec:finitesize}
If we consider a finite size impurity, we need to modify our result within our model as follows. 
We first replace an impurity density by some distribution 
such as the Gaussian distribution, the window function, the Poisson distribution, 
and so on. The Fourier transformation of this distribution is then to be integrated 
over time interval $[t_0,t]$ in Eq.~\eqref{eq:4-2}. For typical distributions 
mentioned above, the integral \eqref{eq:4-2} is written as
\be\label{eq:23}
I_{\v k}(t)=g_\epsilon (\v k) \int_{t_0}^t dt' \Exp{\I \omega_k t'-\I\inn{k}{\zeta(t')}},
\ee
where $g_\epsilon (\v k)$ is some function depending only on $\v k$ and 
$\epsilon$ represents a parameter for sharpness of the impurity distribution 
such that $\lim_{\epsilon\to0}g_\epsilon (\v k)=1$ holds in the limit. 
The Gaussian case, for example, reads $g_\epsilon (\v k)=\exp(-\epsilon k^2)$. 

Critical behavior obtained for the point-like impurity case 
is then valid for the finite impurity case as well. The additional effects 
are solely contributed from a particular form of $g_\epsilon (\v k)$. 
If the function $g_\epsilon (\v k)$ is not singular within the integration interval, 
most of results will not change qualitatively. In particular, 
the scaling in time and the value of critical velocity are same. 
Further analysis needs to obtain more quantitative statement 
and this will be reported elsewhere. 

\subsubsection{Consistency with Landau's argument}
In Landau's argument, one can derive a simple 
consequence: If an impurity in a fluid moves slower than the speed of sound, 
there cannot be any excitations created by the impurity motion.  

The simplest derivation of this statement assumes that 
the total momentum, i.e., the fluid plus the impurity, is conserved 
before and after the interaction between the fluid and impurity. 
Consider the situation where the fluid is at rest initially and 
some excitations are created after the interaction with the impurity. 
We then derive a simple inequality by an another assumption 
of non-increasing in the total energy as 
\be \label{eq:24}
\frac12 m_\imp v_\imp^2\ge\frac12 m_\imp(\v v _\imp-\v p/m_\imp)^2+{\cal E}(\v p),
\ee
where $\v v _\imp$ is the initial velocity of impurity, $m_\imp$ is impurity mass, and 
$\v p$ and ${\cal E}(\v p)$ are momentum and energy of the fluid after the interaction. 
With a straightforward calculation, we obtain the following inequality
\be \label{eq:25}
\v v _\imp\!\cdot\! \v p\ge{\cal E}(\v p). 
\ee
Landau's expression \eqref{eq:1} is thus obtained as a minimal condition 
to satisfy the above inequality \cite{comment3}. 

Apparent inconsistency between Landau's argument and our result 
can be resolved if we notice a simple fact: Our model does not satisfy 
assumptions made in Landau's argument. 
One of them is the conservation of the total momentum. 
Our model Hamiltonian (\ref{eq:2},\ref{eq:3}) do not conserve 
the momentum operator for the Bogoliubov model. This is due to the fact that 
we do not take into account any impurity dynamics and the impurity 
moves without any back reaction from the condensate. 
Thus we need to study back reaction to the impurity to satisfy 
the conservation of total momentum and this is left for further investigation. 

Another assumption that is not satisfied is the energy assumption. 
Again, because the impurity moves along a given trajectory, the total 
Hamiltonian does depend on time in general. This is the situation 
in which we are feeding energy into the condensate 
and thus time-dependent process cannot be explained by Landau's argument. 
For our uniform motion of the impurity, one may argue that 
the Galilei transformation with the same speed turns 
the problem into the static one. As discussed before, we take 
the initial condition as finite time $t_0$ in our calculation. 
This prohibits an application of the Galiei transformation. 
If we impose the infinite remote past condition $t_0\to-\infty$
together with $t\to\infty$ limit, 
on the other hand, we can work in a reference frame where 
the impurity is pinned. 
In this case, we again obtain the infinite limit case as in Sec.~\ref{inftime} and 
we do not observe any dynamics in creation of excitations in the BEC.

Having the above considerations, we conclude that 
Landau's argument cannot be a priori applied to our model. 
Hence, there is no surprise if we have finite number of Bogoliubov 
excitations created from impurity motion below Landau's critical 
velocity. The surprise is in fact that Landau's argument 
predicted \textit{critical} behavior of created bogolons
and this happens at the same critical impurity speed. 

\subsection{Linear response analysis} \label{sec:lra}
In this subsection, we briefly discuss the result obtained from the linear response theory. 
We treat the interaction Hamiltonian as a small perturbation the density of the condensate 
fluctuates due to impurity motion. This density fluctuation measured from 
the expectation value without any impurity is expressed in the thermodynamic limit as 
\begin{multline}\label{eq:26} 
\delta \rho(\v x,t)=g_\imp n \int\frac{d^3k}{(2\pi)^3}\; \frac{\epsilon_k}{\hbar\omega_k}\,\\
\times\big[-\I   I_{\v k}(t)\Exp{-\I\omega_k t+\I\inn{k}{x}}
+\I   I_{\v k}^{*}(t)\Exp{\I\omega_k t-\I\inn{k}{x}}\big], 
\end{multline}
with $I_{\v k}(t)$ given by Eq.~\eqref{eq:4-2}. 
In the following, we give the main result only without detail calculations.

For the finite initial condition, or simply set $t_0=0$, the density fluctuation 
does not show any critical behaviors across $v=c_s$. 
Choose some point in space, say $\v x=0$, we apply 
similar asymptotic analysis to get 
\be \label{eq:27}
\delta \rho(\v x=0,t)\simeq 
\frac{g_\imp n}{(2\pi)^2Mc_s^2\xi^3}\frac{1}{\beta}\,t ^{-3},
\ee
for sufficiently large time $t c_s/\xi$. 
This asymptotic behavior which are same for all values of impurity speed 
can also be checked with numerical integration of Eq.~\eqref{eq:26} directly. 

For the standard treatment of linear response theory, 
we impose the initial condition at $t_0\to-\infty$ limit together with 
the adiabatic factor mentioned in \ref{initcond}. 
In this case, we have instead
\begin{multline} \label{eq:28}
\delta \rho(\v x=0,t)=
-\frac{2ng_\imp}{\hbar} \int\frac{d^3k}{(2\pi)^3}\; \frac{\epsilon_k}{\hbar\omega_k}\,\\
\times\big\{  {\cal P}\frac{\cos[\v k\cdot (\v v t-\v x)]}{\omega_k-\inn{k}{v}}
+ \delta(\omega_k-\inn{k}{v})\sin [\v k\cdot (\v v t-\v x)]
\big\},
\end{multline}
where ${\cal P}$ denotes the Cauchy's principal value. 
Appearance of dirac delta function in the second term 
makes the fluctuation sensitive to Landau's critical velocity 
as analyzed in Sec.~\ref{inftime}. 
This can be reflected by imaginary part of 
the Fourier transformed frequency space, 
i.e., $\mathrm{Im} \{ \delta \rho(\v x=0,\omega)\}$. 
However, as discussed earlier in this section, 
the term $\delta(\omega_k-\inn{k}{v})$ only captures 
different behaviors above Landau's critical velocity, 
but does not provide any information about different scaling behaviors 
below, at, and above the critical velocity. 

To conclude the result from linear response analysis, we 
think that the density fluctuation is not enough to 
analyze critical velocity for BECs in agreement with Refs.~\cite{kw2010,watabe10}. 

\subsection{Comparison with previous results}
We now compare our results with some of previously known results in literature. 
We note that there are many different results and some of disagreements among them seem to arise 
because of not making clear distinction between different aspects of the problem, 
model-dependent result, different properties of superfluidity, etc. 
One also needs to be careful when to compare the result obtained for a specific model 
to other models. Therefore, we only discuss the results 
obtained for similar model analyzed in this paper. 

\subsubsection{Comparison with stability analysis}
In Ref.~\cite{kw2010,watabe10}, Kato and Watabe 
analyzed the autocorrelation function of the local density defined by 
\be \label{eq:29}
I(\v x, \omega)=\sum_{\v k\neq0} 
|\bra{\v k} \hat{\psi}^{\dagger} (\v{ x}) \hat{\psi} (\v{ x}) \ket{gs}|^2
\delta(\omega-\omega_{k}), 
\ee
where $\ket{gs}$ is the ground state of the system, i.e., condensate, 
and $\omega_{k}$ denotes the excitation spectrum measured from the ground 
state energy, see Eq.~(1) of Ref.~\cite{kw2010}. 
They analyzed scaling behavior in frequency for sufficiently 
small $\omega$ and observed different scaling behaviors 
for below or at the critical velocity. This regime corresponds 
to large time asymptotic behaviors analyzed in this paper. 
In the rest of subsection, we only highlight differences from their results. 

Importantly, we are investigating the total number of created Bogoliubov 
excitations and hence different quantities. 
Indeed exponential factors obtained in this paper are different from their result. 
Since their main interest is stability of the condensate around the uniform flow, 
they have not analyzed scaling above the critical velocity, 
where the solution is no longer stable. Above the critical velocity, 
the system is of course unstable and 
we have to be careful when extrapolating the result to 
large values of $v\ge c_s$. We, however, think that 
the system remains quasi-stable for modest values 
of $v> c_s$. This is based on our preliminary analysis 
where the depletion of condensate calculated in this model 
is suppressed even above the critical velocity for sufficiently large condensate, 
yet further detail analysis needs to prove it. 

\subsubsection{Comparison with Casimir type force} \label{sec:caimir}
Robert and Pomeau analyzed drag forces due to 
an obstacle placed in the BEC \cite{rp2005,robert05}, 
see also the result for one dimensional systems \cite{sdr2009,ccb2012}. 
The drag force acting on the BEC is defined by 
\be\label{eq:30}
\v F=-\int d^3x \langle\hat{\psi}^{\dagger} (\v{ x})|\big[\v\nabla \rho_\imp(\v x )\big] |\hat{\psi} (\v{ x})\rangle,
\ee
with $\rho_\imp(\v x )$ a stationary distribution of the impurity. 
Based on the generalized Gross-Pitaevskii equation which takes 
into account the impurity localized at some position, that is 
classical condensate wave function is no longer homogeneous, 
they showed that a finite drag force exists at arbitrary small velocity of flow. 

The common feature between their result and ours is that 
we have measurable effects even below the critical velocity. 
Note that creation of excitations does not immediately imply friction, 
that is our result does not tell if there exists finite force acting on the impurity or not. 

Disagreement is, on the other hand, their statement about 
interpreting the calculated force as Casimir force. They explicitly 
stated that there should be no excitations created below Landau's 
critical velocity, yet finite amount of force arises due to what they called 
``scattering of zero-temperature quantum fluctuations" \cite{rp2005}, 
like as in the ordinary Casimir effect. We suspect without any detail calculations 
that their obtained force is partially due to the interaction with 
excitations with low momenta since they do have finite number of excitations 
below the speed of sound. 

\subsubsection{Fermi's golden rule for finite time}
Lastly, we comment of the usage of Fermi's golden rule 
to calculate the decay rate, which was mentioned in Sec.~\ref{inftime}. 
Recently, Fermi's golden rule for finite time was analyzed by 
Ishikawa and Tobita \cite{it2013}. They analyzed 
scattering processes without taking infinitely remote past and future limit. 
They also pointed out some of physics will be changed 
whether we take $t\to\infty$ limit first or large wave packet size limit. 
Although their analysis is totally different, it may be interesting 
to see their modified Fermi's golden rule with finite time corrections 
provides a similar conclusion as obtained in this paper. 
Another interesting question would be to study relativistic 
Cherenkov radiation with our finite time analysis and 
to ask if there exists a similar critical behavior. 

%

\section{Summary and outlook} \label{sec:5}
In this paper, we have confirmed that there can be a finite number of Bogoliubov 
excitations created in the BEC from a uniform impurity motion even below Landau's 
critical velocity, which is the speed of sound. Importantly, our result was obtained 
for the homogenous three-dimensional condensate at zero temperature. 
Our result suggests that one should not count 
the number of excitations created as a function of impurity speed 
as opposed to previous experimental results \cite{becExps1,becExps2,becExps3,becExps4,becExps5}. 
In this case, one can never observe critical behavior for finite time experiments 
around this critical velocity, see FIGs.~\ref{fig1}, \ref{fig2}. 
In this paper, we have proposed 
a different way of observing critical behaviors, that is to measure the number of 
excitations as a function of time interval for which the impurity travels. 
This kind of experiments is expected to show 
completely different scaling behaviors for below, at, and above Landau's critical velocity. 
In summary, the transient excitations due to the impurity motion reveal more transparent 
meaning for Landau's critical velocity rather than looking at the steady-state excitation rate. 

It is important to examine whether we can observe different scaling behaviors for 
other types of trajectories, such as rotations with a constant angular velocity or 
dipole-like oscillation with a constant frequency. 
Based on a similar analysis presented in this paper, 
our preliminary investigation shows indeed critical behaviors depending on 
its ``speed." Because the speed of the impurity is not uniquely defined 
by a single time-independent quantity 
for general motions other than the uniform motion, the notion of speed 
needs to be clarified or to find critical values for other quantities. 
This matter will be presented in future publication separately. 

One question we have not addressed so far is a relationship to the onset of superfluidity. 
From our result, different scaling behaviors might be used as one of the measure to detect superfluidity in one sense, 
that is there are only constant amount of excitations in the BEC for infinite time limit 
if the impurity speed is below the critical speed. 
However, the presence of constant number of excitations does not immediately imply frictionless motion for the impurity, 
in particular, this is because the constant obtained in this paper, 
${\cal N}_0(v/c_s)^{-1}\log[(c_s+v)/(c_s-v)]$, is a function of the impurity speed. 
Thus, it is necessary to analyze back reaction to the impurity 
in order to check the onset of frictionless motions, which is more directly 
related to the phenomenon of superfluidity. This deserves to further investigation and 
we plan to analyze it in due course.


\begin{acknowledgments}
The author wishes to thank Cord A.~M\"uller for many valuable discussions and constructive comments on the paper. 
He thanks Shohei Watabe for providing useful information regarding Refs.~\cite{kw2010,watabe10} 
and stimulating discussions about this subject. 
He also acknowledges Berge Englert and Christian Miniatura for helpful discussions 
and their kind hospitality at Centre for Quantum Technologies where part of this work was done. 
\end{acknowledgments}


\begin{thebibliography}{99}
\bibitem{comment1}
The rate, defined by a time derivative of Eq.~\eqref{eq:5-1}, 
provides the same expression for large time limit upon the usage of 
another formula $\lim_{t\to\infty} \sin\omega t/\omega=\pi\delta(\omega)$. 
An implicit assumption is that commutativity among the time derivative, the limit, 
and the integration over $k $.
\bibitem{comment0} 
Radiation in a solid angle $d\Omega$ is defined by 
$\int dk k^2 \hbar \omega_k n_{\v k}(t)$. The total amount of radiation is then 
obtained by integrating over angles.
\bibitem{comment3}
When an impurity interacts with a fluid, the mass of impurity may be dressed 
due to the interaction with the fluid. In this case the inequality 
\eqref{eq:25} is no longer valid and hence Landau's inequality also needs 
to be modified. This mass effect was already discussed by G. Baym in Ref.~\cite{baym}.
\end{thebibliography}
\end{document}